# Measuring Cognitive Activities in Software Engineering

**Pierre N. Robillard**
**Patrick d'Astous**
École Polytechnique de Montréal
C.P. 6079, Suc. Centre-ville
Montréal , CANADA  H3C 3A7
514-340-4238
robillard@rgl.polymtl.ca
dastous@rgl.polymtl.ca

**Françoise Détienne**
**Willemien Visser**
Ergonomic Psychology Project
INRIA, Rocquencourt
B.P. 105, 78153 Le Chesnay, FRANCE
francoise.detienne@inria.fr
willemien.visser@inria.fr

**ABSTRACT**
This paper presents an approach to the study of cognitive activities in collaborative software development.  This approach has been developed by a multidisciplinary team made up of software engineers and cognitive psychologists. The basis of this approach is to improve our understanding of software development by observing professionals at work. The goal is to derive lines of conduct or good practices based on observations and analyses of the processes that are naturally used by software engineers.  The strategy involved is derived from a standard approach in cognitive science. It is based on the videotaping of the activities of software engineers,  transcription of the videos, coding of the transcription,  defining categories from the coded episodes and  defining cognitive behaviors or dialogs from the categories. This project presents two original contributions that make this approach generic in software engineering.  The first contribution is the introduction of a formal hierarchical coding scheme, which will enable comparison of various types of observations. The second is the merging of psychological and statistical analysis approaches to build a cognitive model.   The details of this new approach are illustrated with the initial data obtained from the analysis of technical review meetings.

# 1 INTRODUCTION
Measurement is an important component of engineering activities, and software engineering is no exception.  Measurement can be useful for assessing the software production process, the intrinsic characteristics of software and the utility of software systems in their environment [1].

The software system in its environment is often measured in terms of the software reliability where software errors and system failures are the basic factors of most measurements.  The intrinsic characteristics of software are usually related to its measurements, based on source code, or to its derivatives, and are considered part of the software complexity domain. The software production process is measured in terms of time spent in various activities and the volume of the resulting artifacts, either code or documentation, and it composes the software productivity domain. Software development is knowledge-intensive [2], however the cognitive activities involved in software development have been studied little so far.  This should not be mistaken with the domain of the psychology of programming, which is mostly interested in the use of programming languages. This may be explained by the fact that the cognitive science is not part of the curriculum for the software scientists or engineers. In Human Computer Interaction (HCI), a domain where cognitive sciences are involved, the study of human behaviors has proven to be rewarding. The domain of study of human behavior in software engineering has been called " psychology of programming ".
Cognitive psychologists have developed many approaches to the study of mental activities [3]. The cognitive activities in software engineering can be studied from the individual or the team perspective.  The individual aspect of a cognitive activity, which has been so far the subject of most of the empirical studies,  is concerned especially with the mental mechanisms, the strategies and the knowledge used to perform the activity.  For example, the way a designer uses the concept of object to create a design. The cognitive process involved in team interactions is the team aspect; for example, the cognitive process involved in a technical review meeting.

This paper presents an approach to the study of the cognitive activities involved in team work. A multi-disciplinary team made up of software engineers and cognitive psychologists has developed this approach.

The basis of this approach is the desire to improve our understanding of software development through the observation of software professionals at work. The goal is to derive modes of conduct or good practices. The strategy was inspired by research practices in cognitive science. It is based on the videotaping recording of the activities of software engineers, transcription of the videos, coding of the transcription, thorough analysis of the codes and modeling of cognitive behaviors from the analyses. This project presents two original contributions that make this approach generic in software engineering. The first contribution is the introduction of a formal hierarchical coding scheme in order to enable comparison of various types of observations. The second contribution is a complementary approach to the derivations of dialog types: by cognitive psychologists on the basis of their expertise in modeling cognitive activities and in a formal way by statistical analysis of the coded transcripts. The two analyses are then confronted in order to build a cognitive model.

This approach has been applied to one type of meeting in a professional software development project based on a defined software development process. The project required 4 full-time software engineers and lasted 19 weeks.

A software-development project involves various types of meetings where team members exchange ideas, review the work done or plan future tasks. These meetings can be formal or informal and of various types such as the design review, the technical review, the walkthrough and the code inspection.

The technical review meeting (TRM) has been selected to implement this new approach. To date, cognitive analyses have been performed on design meetings [4] and code inspection meetings [5]. A TRM has two main objectives: to verify the current state of the design project and to validate the specifications of the succeeding tasks. A TRM requires the presence of several reviewers for a certain amount of time (in the order of thirty minutes to two hours). TRMs are held throughout the development process to ensure that the specifications of the software components have been met and that the tasks have been carried out according to the recommended practices. A project team can hold various TRMs during a week. Various references have outlined the activities that are supposed to take place in a technical review meeting [6] [7].

The contribution of TRM analysis is twofold. In Cognitive Psychology, it will increase the knowledge of the collective activities involved in software development. In Software Engineering, it will provide the necessary knowledge to assess the current guidelines and to render them more suitable to the practitioner's needs.

The following section outlines the difficulties and the corresponding solutions for this measurement approach. The next section introduces the coding scheme and shows the formalized notation proposed. Finally, the derivation of dialogs and some sample analyses are presented.

This work stresses the benefit of collaboration with cognitive psychologists to introduce new links in software engineering. The results presented in this paper illustrate the potential of this measurement approach.

Software engineering processes are made up of a set of practices, and these practices are carried out by human beings using cognitive activities. The goal of this approach is to study these cognitive activities in order to improve the corresponding software practices. This paper illustrates their application to the study of practices surrounding the Technical Review Meeting, but this approach could be applied to any practice.

## 2 MEASUREMENT APPROACH

The observational approach used was to videotape technical review meetings, because it does not disturb the meetings, and thus allows valid data to be collected.

A specially trained typist then transcribed the videos. Each individual intervention is a transcript entry. A good transcript is not trivial to establish and many trials were needed before the right way to do it was found. The main difficulties arise from the naturalness of a meeting. After a time, people forget that they are being observed by a camera and they begin to use familiar expressions, very technical words, talk at the same time, interrupt and tell jokes; in short, exhibit very natural behavior which makes a meeting interesting but its transcript arid. These transcripts form the basic data for the analysis, although it is sometimes useful to refer back to the video to validate the meanings of some sentences.

The transcripts were transformed by the following means:
- defining categories based on a coding scheme that describes the activities occurring in a review meeting;
- restructure individual interventions into episodes using a decomposition rule.

A coding scheme has been developed, which enables the coding of episodes using syntactically structured labels, also called categories. Individual interventions have been decomposed into episodes using the following decomposition rule: a new episode starts when the activity in the intervention can no longer be covered by the same category. A new formal coding scheme was developed for this project. This coding scheme must meet the theoretical criteria of interest and be objective [8]. Objectivity deals with the reliability and validity of the coding scheme, while theoretical interest depends mainly on the domain and on the goal of the research. Categories must be exhaustive and exclusive. A code must be able to model the meeting activities adequately and yet be formal enough to support quantitative analysis. Figure 1 presents the generic steps in this measurement approach.

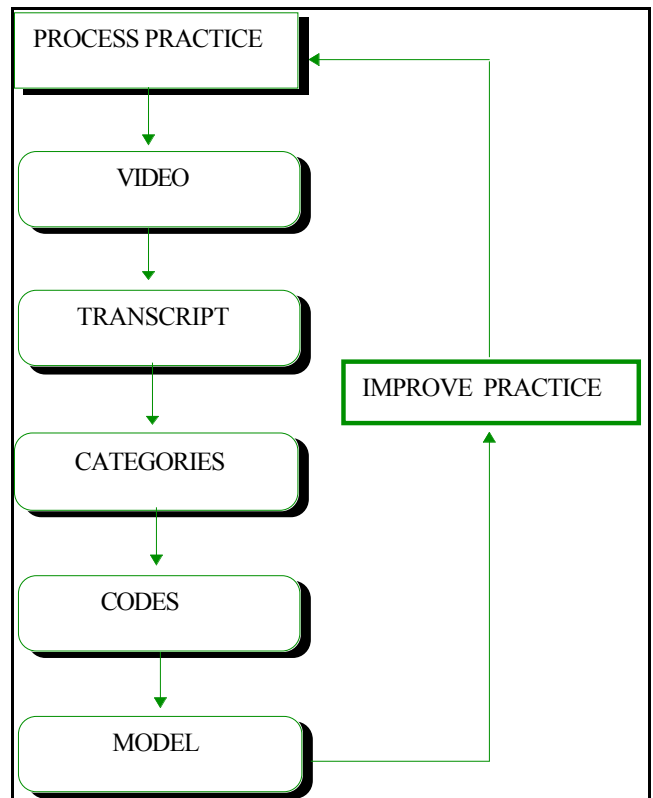

**Figure 1 Measurement approach**

Most experiments that have dealt with the observation of participant behavior in the software process were carried out on design meetings. Olson et al. developed a coding scheme, which contained 22 activity categories depicting the general nature of the design discussions [9]. The coding scheme was then used to observe the interactions of a group of experienced software designers during design meetings [10]. They found that a great deal of time was spent on design discussions involving the design objects and their clarification.

Herbsleb et al. [11] used the same coding scheme to compare design meetings in both the procedural and object-oriented environments. Their goal was to compare the team aspects of procedural and OO software development in order to evaluate the claims made for the superiority of object orientation. They obtained results very similar to the previous results with respect to the allocation of time in the three main categories.

Another approach to the study of meetings was used by Karsenty [12] which was to observe participants validating the conceptual schema of a database. The coding scheme used in this experiment was based on four types of dialogue (evaluation, clarification, negotiation and problem analysis). The results obtained from its use enabled the author to measure the importance of the schema in representing a solution to future users and to contain the specifications required to construct the database.

These coding schemes were studied for this project. We concluded that none was sufficient to account for the cognitive activities involved in TRMs. Therefore, a new coding scheme was developed, by adding categories and, progressively, in an iterative way, refining the existing ones. The result is an applied and validated formal hierarchical coding scheme. [13]

Categories are then used to build dialogs of the activities, and a cognitive model is developed to analyze this dialog.

The resulting analysis hopefully provides a better understanding of the TRM process practices studied. Improvements of these practices will be proposed and the cycle may start over until optimized practices are obtained.

## 3 CODING SCHEME

A three-level coding scheme was designed to facilitate easy identification of the episodes and to provide the basis for coherent analysis:
1. Level one identifies the type of activities performed by the reviewer,
2. Level two defines the entity on which an activity is performed and,
3. Level three, which is optional, is the criterion that complements the entity.

For example, the teammates can discuss (activity) the design document version x (entity) and the point of discussion refers to the format used to describe the design (criteria of form).

Although the structure of this coding scheme is general, it needs to be customized for a technical review meeting by specifying the type of activities, the nature of entities, and the type of criteria likely to be found in such meetings.

The customized coding scheme is formed of four distinctive groups of activities: reading, discussing, requesting and managing. Progression through a TRM is based on the reading of each individual section (which are used as a labeled reference) of the document being revised (artifact). Each of these readings can bring forth one of the many discussion activities. The discussing activity is the main activity of the TRM and it operates on objects. The objects on which these discussions are performed can either be an artifact, which is a written document or the outcome of a previous cognitive activity, called "message". The objects can be further specified by a criterion of form or content. Criteria are based on the content or on the form of the object. Content criteria are derived, for this project, from the quality attributes of the ISO 9126 standard. Form criteria are derived from the programming and the documentation guidelines. Table 1 illustrates the customization of the generic coding scheme for TRMs. Any participants to bring new ideas or alternative solutions can also present an alternative statement.

The *request* activity is initiated by a reviewer on an object that could be an artifact or a message. The *manage* activity is performed on tasks that may be at the project level, for example planning the next review, or at the meeting level, for example when the moderator suggests that a specific section of the document be reviewed.

| ACTIVITIES | ENTITIES | CRITERIA |
|---|---|---|
| *manage* | *task* | |
| *read* | *artifact* | |
| *request* | *object* | |
| *discuss* | *object* | *form or content* |

**Table 1 Customization of the coding scheme for TRM**

The *discuss* activity is generic, and can be further specified by one of the eight detailed activities. Some of these detailed activities can be even further defined if required by the analysis. For example, the *evaluate* activities could be specified as either positive or negative and the *hypothesize* activities can specify the purpose of the hypothesis, such as to better explain, to justify or to inform.. The definitions of the cognitive activities that compose the discussion and their resulting messages are presented in Table 2.

Figure 2 presents the formal representation of the coding scheme based on a BNF description [14]. A coded episode is made up of at least one of the four optional items available in the coding scheme. Each episode is defined by an activity, a separator symbol (/) and an appropriate entity.

A criterion can be added in the case of the *discuss* activity.

The following two examples show the use of artifacts and messages respectively as objects in a coding of episodes.

| Activity/ Message | Abr | Definition |
|---|---|---|
| **Accept/** | ACC | Validating a particular |
| *Acceptance* | | object. |
| **Develop/** *Development* | DEV | Elaborating a new concept for an existing object or criterion. |
| **Evaluate/** *Evaluation* | EVAL | Analyzing an object according to a criterion. Can either be positive or negative. |
| **Explain/** *Explanation* | EXPL | Providing information on the HOW of an object or criterion. |
| **Hypothesize/** *Hypothesis* | HYP | Representing the HOW, WHAT or WHY of an object or criterion. |
| **Inform/** *Information* | INF | Handing out information with respect to the nature (WHAT) of an object or criterion. |
| **Justify/** *Justification* | JUST | Arguing or proving the rightness of WHY a particular choice was made. |
| **Reject/** *Rejection* | REJ | Discarding an object as invalid. |

**Table 2 Discussion activities and corresponding messages**

```
Symbol definitions      :
            /                       separator
            ::=                     meaning "is defined as"
            |                       meaning "exclusive or"
            <>                      category names
            {} +                    at least one item must exist
            {} 0                    the item is optional

< code   >::=    {{manage / <task>}        0 | {read / <artifact>}        0 |
                 {request / <subject>}     0 |
                 {<discuss> / <subject> {/ <criterion>}      + } 0 } +

<discuss>   ::=  accept | develop | evaluate | explain |
                 hypothesize | inform | justify | reject
<task>      ::=  project | meeting
<subject>   ::=  <artifact> | <message>
< artifact >::=  document | document section
< message  >::=  acceptation | development | evaluation |
                 explanation | hypothesis | information |
                 justification | rejection
< criterion  >::=  form | content
```

**Figure 2 BNF representation of the generic coding scheme customized for technical review meetings**

An episode that records the evaluation of a reviewer regarding section *n* of the reviewed document based on a criterion from the programming guide is coded as:
> *EVALUATE/SECTION-n/CRITERION*

In the following episode, where another reviewer rejects the evaluation just made, the episode would be coded as
> *REJECT/EVALUATION-m/CRITERION*

The letter *m* is a label that uniquely defines this message. All the episodes that compose a review meeting can be coded in this way using all the reviewing activities, entities and criteria. The sequential list of coded episodes is a context-free representation of the meeting.

This coding scheme enables the translation of the transcript into a list of coded episodes. The generic approach to coding and the use of the activity/entity/criterion structure for every type of activity make the coding of episodes easier and more uniform.

On the basis of the attribution of different categories to the episodes, dialog types have been defined by identifying sequential patterns of coded episodes. . The patterns can be derived using the psychological approach or the statistical approach. The data presented in this paper are mostly based on the psychological approach, which requires expert psychologists to study the categories describing the meeting and derive dialog types on the basis of category patterns. This approach produces the following five dialog types:

1. Review (REV) dialog is the *raison d'être* of the technical review meeting. It is characterized by cycles of evaluating - justifying activities
2. Alternative elaboration (ALT) dialog arises when a reviewer proposes a solution that is not described in the review document. This dialog is dominated by the develop activities.
3. Cognitive synchronization (SYNC) [15] dialog enables the participants to insure that they share a common representation of the state of (alternative) design solutions or of evaluation criteria (content or formal). This dialog is characterized by request-inform-hypothesize activities
4. Conflict resolution (CONFL) dialog results when reviewers do not agree on what is being discussed. There is argumentation between two or more participants regarding a conflict generated by diverging opinions on criteria or (alternative) solution evaluation, or by diverging representations of the state of the (alternative) design solutions or of evaluation criteria. This dialog is based on reject - evaluate - justify activities
5. Management dialog is required for the coordination and planning of different tasks

## 4 DATA SAMPLES

This section presents initial data that illustrate the appropriateness of this measurement approach to the study of meeting activities. The data are obtained from the analysis of a technical review meeting performed by four individuals, which lasted 25 minutes and contained 256 coded episodes. Coded protocols can be analyzed from three viewpoints: frequency distribution and time distribution of episodes within categories, and episodes grouped into dialog patterns.

These data were taken from the observation of one technical review meeting and cannot realistically be used as a basis for a valid argumentation, but they show quite well the extent to which such measurement tools can be used.

### 4.1 Frequency and Time Distribution of Categories

The frequency distribution shows the relative importance of coded episodes and it is an indicator of the suitability of the customized coding scheme. For example, if a category is strongly dominant, which means that most of the episodes are coded in the same category, then the coding scheme presents a weak sensitivity to this type of activity. Figure 3 shows the relative frequency importance of the managing (MNG), reading (READ), requesting (RQST) and discussing (DCSS) activities.

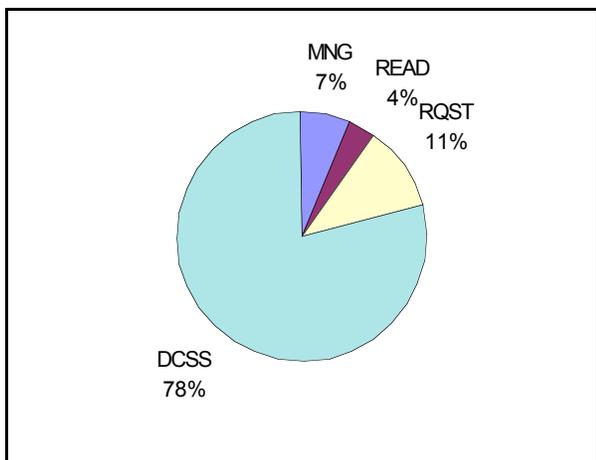

**Figure 3 Frequency distribution of activities**

The *discuss* activity is dominant, and has therefore been subdivided into 8 more detailed activities. Figure 4 shows the frequency distribution of these detailed activities within the *discuss* activity. Most of the activities are evenly distributed except for the *reject* (REJ) activities. Considering the frequency occurrence of the categories it is possible to validate the sensitivity and appropriateness of this coding scheme for technical review meetings.

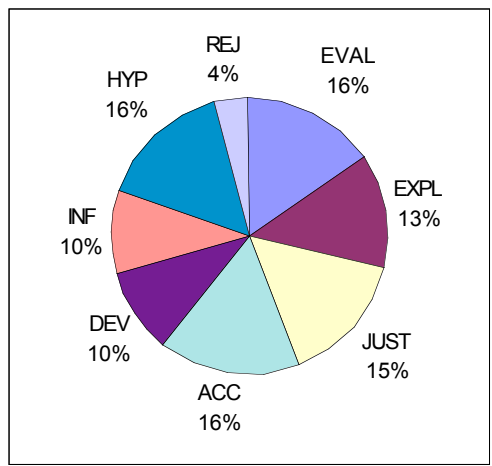

**Figure 4 Frequency distribution of activities within the discuss category**

The analysis of time category distributions in Figure 5 shows how time is consumed by the different activities. The *read* activity illustrates the relationship between the frequency and time viewpoints. Although the frequency of the *read* activity is not significant (< 5%), the time spent in the *read* activity is (>15%).

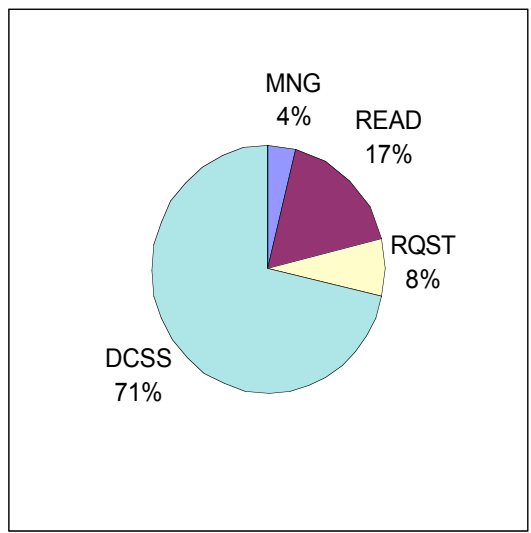

**Figure 5 Relative time distribution of activities**

Figure 6 shows that the *explain*, *develop* and *justify* activities take most of the time of the review meeting. The *accept* activity, which is the purpose of the review meeting, occupies only 5% of the time spent in discussion. However, this activity accounts for 16% of the frequencies of occurrences. This means that the episodes resulting in the *accept* activity have a short time duration.

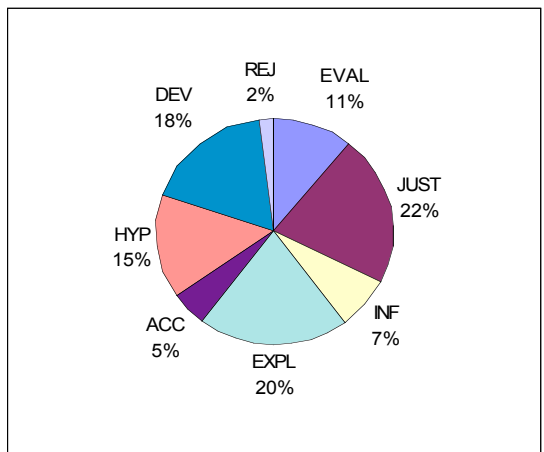

**Figure 6 Relative time distribution of activities within the discuss category**

Analysis can be performed on the distribution of the meeting time among the objects. Figure 7 shows the time spent in discussing the initial solution (INI. SOL) of the reviewed document and the alternative solutions (ALT. SOL.) spawned by those discussions, in evaluating criteria (CRIT) and finally, other issues (OTH).

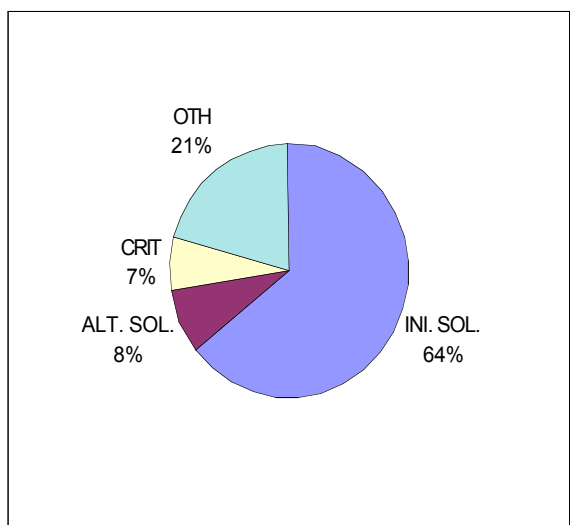

**Figure 7 Relative time distribution of objects**

Even though a typical technical review meeting is devoted to the review of a given document, Figure 7 clearly shows that more than 35% of the meeting time is spent in discussing other issues.
The characteristics of a technical review meeting can be defined from the analysis of the various activities, entities and criteria.

**4.2 Patterns of Dialogs**

The dialogs are intertwined and careful analyses are required to identify them correctly. Figure 8 shows the relative time distribution of the three dominant dialogs. Half of the time is occupied by the cognitive synchronization (SYNC) dialogs and the other half is spent equally between the review (REV) dialog and the alternative elaboration (ALT) dialogs.

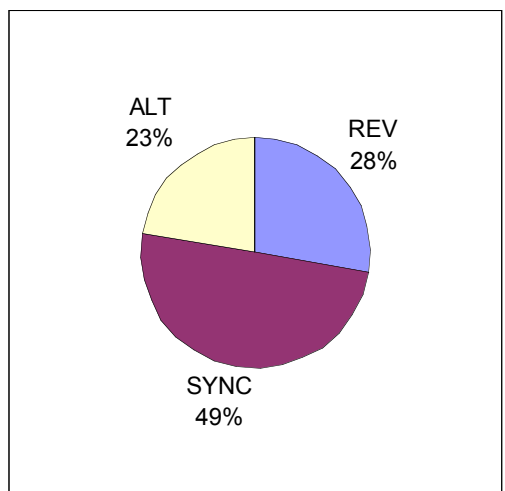

**Figure 8 Relative time distribution of the dialogs**

This pie chart does not include the conflict resolution (CONFL) dialog because this is really a sub-dialog which usually occurs within an existing dialog. Figure 9 shows that conflict resolution dialog accounts for at least 10 % of the activities and for up to 20% for cognitive synchronization dialog.

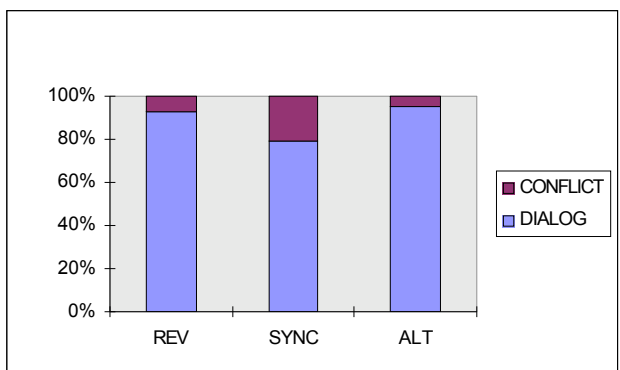

**Figure 9 Relative time distribution of conflict resolution within dialogs**

Figure 10 shows the dialog time duration for each of the 12 sections of the document under review. The evaluation dialogs are usually short and take less than a minute. Some sections need extensive cognitive synchronization dialogs, and alternative solution dialogs are then initiated.

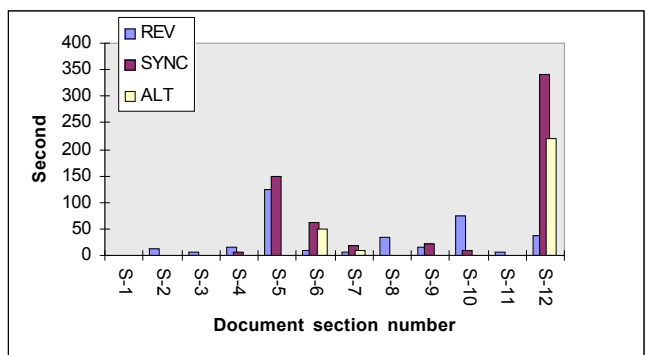

**Figure 10 Dialog time duration for each section**

### 4.3 Statistical Analysis of Categories

Statistical analysis of the categories can also be used to derive various types of dialogs. Statistical tools are required to find the categories that are forming a sequence with an acceptable level of confidence. The use of probabilistic techniques and log-linear modeling will facilitate the observation of patterns from the initial categories. Such patterns are then analyzed to understand what makes them occur or not.

Modeling is the estimation of parameters describing a situation from collected data. Linear models, which combine expected values in a linear fashion, are used extensively for such work. The adjustment of a log-linear model requires the validation of hypotheses made on the data through the estimation frequencies and by comparing these frequencies to the observed values. This type of modeling enables the observation of patterns in the data that will lead to the validation of dialogues. The initial data presented in this analysis are insufficient to obtain significant results from the use of log-linear modeling.

Another approach is to determine the sequence of categories that are more significant than random activities. The analysis of sequential structures comes from information theory. Lag Sequential Analysis (LSA) enables the identification of categories that follow one another, even with other categories in-between. The analysis consists in determining whether or not the frequency of a given category is independent of the frequency of another category. Sequential structures enable the definition of patterns. This technique requires less data in order to be significant.

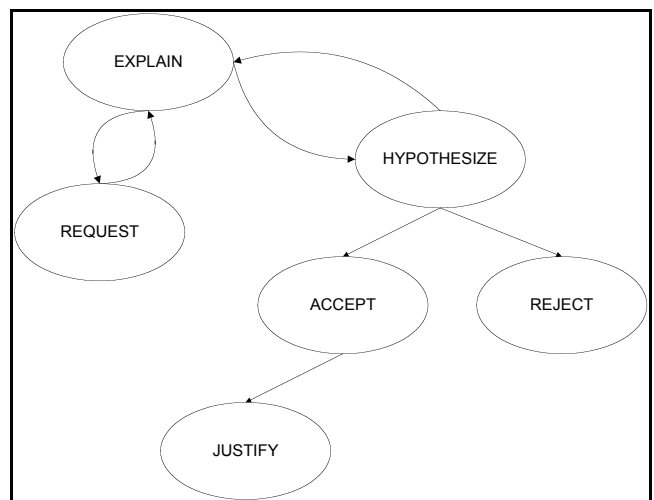

**Figure 11 Identified pattern using LSA**

Figure 11 shows an example of the identification of a pattern in the data using LSA. This pattern can be related to the cognitive synchronization dialog identified earlier.

This approach can validate the pattern found by the psychologist and the psychologist can help in interpreting and validating the patterns that are identified by the statistical approach.

### 5. CONCLUDING REMARKS

Software engineering processes rely heavily on the ability of software engineers to successfully accomplish the required tasks. Measurement of the software product itself and the software in its environment has lead to better engineering practices. Measuring the software activity in its actual day-to-day work context should contribute tremendously to the improvement of software process practices.

The measurement approach and the coding scheme presented in this paper will provide a systematic procedure to facilitate the observation and analysis of software engineers at work. Formal and informal communications form the raw data on which this technique is based because communication is the fundamental element of team interaction. Understanding and facilitating it will improve software engineering practices.

This paper presents the use of the measurement approach on technical review meetings because they have a very important but often neglected role in the software process. As a formal mean of communication, they must be dependable and efficient. Their measurement and resulting conclusions will be a step in this direction.

For example, this study shows that technical review meetings seem actually to be composed of three types of cognitive activities: review, synchronization and elaboration of alternative solutions. Simulation and further research are needed to build and validate a procedure for technical review meetings where the cognitive synchronization activities will be minimized. Maybe a new type of meeting dedicated to the cognitive synchronization will greatly improve the software engineering process.

A better understanding of the intrinsic characteristics of meetings, and in particular the review meeting, will provide the necessary knowledge to assess the current practices or to render them more suitable to the practitioner's needs. This cannot be done without a thorough investigation of actual behaviors.

This paper illustrates an approach to rigorously measuring team behavior during meeting. This approach is general and can be applied to any type of meeting. Customization involves refining a generic coding scheme by incorporating specific activities, entity and criteria.

## ACKNOWLEDGMENTS

Special thanks are due to Jim Herbsleb for helping us in understanding the requirements of such a research project.

The analysis and study required to create and validate the coding scheme could not have been carried out without the technical help of Benoit Lefebvre and Luc Lalande of the Software Engineering Lab at the Ecole Polytechnique in Montreal.

This project was supported in part by the STEP program, le programme d'échange France-Québec and NSERC grant A0141.